\documentclass{aa}
\usepackage{amsmath}
\usepackage{amssymb}
\usepackage{graphicx}
\usepackage{natbib}
\bibpunct{(}{)}{;}{a}{}{,}

\begin{document}
\title{Tidal effects on small bodies by massive black holes}
\author{
				Uro\v s Kosti\' c
				\inst{1}
				\and
				Andrej \v Cade\v z
				\inst{1}
				\and
				Massimo Calvani
				\inst{2}
				\and
				Andreja Gomboc
				\inst{1}
				}
\offprints{U. Kosti\' c, \email{uros.kostic@fmf.uni-lj.si}}
\institute{Faculty of Mathematics and Physics, University of Ljubljana, Jadranska 19, 1000 Ljubljana, Slovenia
					\and
					INAF - Astronomical Observatory of Padova, Vicolo Osservatorio 5, 35122 Padova, Italy
					}
\date{Received ?? / Accepted ??}
\abstract
{The compact radio source Sagittarius $\mathrm{A}^*$ (\mbox{Sgr~$\mathrm{A}^*$}) at the centre of our Galaxy harbours a supermassive black hole, whose mass ($\approx 3.7 \times 10^6 M_{\sun}$) has been measured from stellar orbital motions.
\mbox{Sgr~$\mathrm{A}^*$}\ is therefore the nearest laboratory where super-massive black hole astrophysics can be tested, and the environment of black holes can be investigated. Since it is not an active galactic nucleus, it also offers the possibility of observing the capture of small objects that may orbit the central black hole.}
{We study the effects of the strong gravitational field of the black hole on small objects, such as a comet or an asteroid. We also explore the idea that the flares detected in \mbox{Sgr~$\mathrm{A}^*$}\ might be produced by the final accretion of single, dense objects with mass of the order of $10^{20}\ \mathrm{g}$, and that their timing is not a characteristic of the sources, but rather of the space-time of the central galactic black hole in which they are moving.}
{The problem of tidal disruption of small objects by a black hole is studied numerically, using ray-tracing techniques, in a Schwarzschild background.}
{We find that tidal effects are strong enough to melt sufficiently massive, solid objects, and present calculations of the temporal evolution of the light curve of infalling objects as a function of various parameters. Our modelling of tidal disruption suggests that during tidal squeezing, the conditions for synchrotron radiation can be met. We show that the light curve of a flare can be deduced from dynamical properties of geodesic orbits around  black holes and that it  depends only weakly on the physical properties of the source.}
{}
\titlerunning{Black hole's tidal effects}
\authorrunning{Kosti\' c et al.}
\maketitle
\section{Introduction}
The centre of our Galaxy harbours the nearest massive black hole. Stellar orbits determinations \citep{2005ApJ...620..744G} revealed a central dark mass of $(3.7\pm 0.2)\times 10^6 [R_0/(8\, {\rm kpc})]^3 M_\odot$, with an updated value of $(3.61\pm 0.32) \times 10^6 M_\odot$ from observations with SINFONI \citep{2005ApJ...628..246E} confined within a radius of 45 AU. The proximity of the Galactic centre ($\approx 8$ kpc) allows us to study the environment of massive black holes in  detail \citep[see e.g.][]{2005PhR...419...65A}.

Although the black hole at the Galactic centre is not an active one, flares from its direction have been detected in X-rays (Chandra and XMM-Newton satellites) and infra-red (VLT adaptive optics imager NACO, SINFONI infra-red adaptive-optics integral spectrometer on the ESO VLT). These flares are modulated on a short timescale with average periods of $\approx 20$ minutes \citep{2003Natur.425..934G, 2006JPhCS..54..420B, 2006A&A...455....1E, 2006A&A...460...15M}. Multi-wavelength campaigns found that the time lag between X-ray and NIR flare emission is small, strongly suggesting a common physical origin \citep{2004A&A...427....1E, 2006A&A...450..535E, 2006ApJ...644..198Y, 2007MNRAS.375..764T}.

Several different models were proposed to explain these phenomena: disk instabilities (e.g.\  \citet{2006ApJ...636L..33T}, \citet{2006JPhCS..54..427Y}, \citet{2008ApJ...679L..93F}, \citet{2008A&A...479..625E}), star-disk interaction (e.g.\  \citet{2004A&A...413..173N}), expanding hot blobs (e.g.\  \citet{2006ApJ...644..198Y}, \citet{2008ApJ...682..373M}), hot spot/ring models (e.g.\  \citet{2007MNRAS.375..764T}, \citet{2004A&A...427....1E}, \citet{2006A&A...450..535E}, \citet{2006JPhCS..54..420B},  \citet{2006A&A...460...15M}, \citet{2006A&A...458L..25M}, \citet{2006MNRAS.367..905B}, \citet{2008JPhCS.131a2008Z}).

The article by \citet{2005PhR...419...65A} contains an exhaustive description of the Galactic centre (GC) environment and its stellar dynamics. It is reasonable to expect that stars at the GC are surrounded by planets and low-mass satellites, e.g.\  comets and asteroids. If the population of these objects is similar to that around the Sun, there must be a considerable number of solid objects that move around the GC. In our previous paper (\citet{2008A&A...487..527A}, hereafter \v CCK08), it has been shown that tidal evolution of orbits can bring their periastron close to the black hole. If an object finds itself below its Roche radius, tidal forces dominate and disrupt it. In this paper, we study the possibility that tidal disruption events are connected with the observed flares from the GC. The timescales of flares strongly suggest motion only a few Schwarzschild radii away from the black hole and it has been speculated that the energy released during these flares corresponds to the source mass of the order of $\sim 10^{20}\ \mathrm{g}$ \citep{2003Natur.425..934G}. Asteroid-like objects explain both observational results: they are of the correct mass and are tidally disrupted closer to the black hole than gaseous blobs. In Active Galactic Nuclei, these events would be undetectable, but in the inactive centre of our Galaxy, it is possible to observe them.

In Sect.\hspace{2pt}\ref{sec:small}, we review some ideas regarding the population of these small bodies in the immediate neighbourhood of the Galactic centre black hole and discuss the difference in the Roche radius for solid and gravity-dominated objects. In Sect.\hspace{2pt}\ref{sec:tidal}, we study tidal disruption of a small object numerically, using ray-tracing techniques in a Schwarzschild background. We present calculations of the temporal evolution of the light curve of infalling objects as a function of various parameters. In Sect.\hspace{2pt}\ref{sec:discussion}, the possible consequences of tidal squeezing on the object's internal magnetic field are discussed, and calculated light curves are compared with the observed light curves of flares in the Galactic centre.

\section{Small bodies at the Galactic center}
\label{sec:small}
It appears reasonable to assume that stars at the GC have planets and other small orbiting bodies, such as asteroids and comets. When the parent star approaches the central black hole, tidal interaction may either strip these bodies off their parent star or cause them to become more bound. In this way, the star may lose its satellites before it is swallowed or disrupted by the black hole. Little is known about the dynamics that determine the fate of these satellites.

A related problem was studied by \citet{2006ApJ...650..901B}, who investigated cluster core dynamics in the GC, by considering intermediate massive black holes that transport multiple tightly-bound stars inwards into the potential well to a tidal-stripping radius. They found that this mechanism is likely to produce a population of stars close to the GC on high-eccentricity, high-inclination orbits. An analogous mechanism could be applicable to a planetary systems of stars.

The possibility of detecting gravitational waves with LISA stimulated an extensive investigation of mechanisms to cause stars to move inwards into a massive black hole \citep[and references therein]{Hopman2006}. These studies demonstrated that the central density cusp of the stellar distribution flattens to $n(r)\propto r^{-1/2}$ at a critical radius of $a_{\rm GW}=\rm{few}\times0.01\ \rm{pc}$ for the distribution of $1\ M_\odot$ objects. Here $a_{\rm GW}$ was defined to be the radius where gravitational radiation damping prevails over resonant relaxation in the central star cluster. Since the power of gravitational wave emission is proportional to $m^2$, $a_{\rm GW}$ is negligible for small bodies. Consequently, their central density cusp may extend much further to radii at which energy and/or angular momentum can be extracted efficiently from the orbit by tidal interaction, as discussed by \v CCK08.

A beautiful presentation of the Galactic centre environment based on high resolution observations is given by \citet{2006ApJ...643.1011P}. They  confirm a marginally steeper stellar density cusp $n(r)\propto r^{-2}$, and its flattening, in fact a sharp drop, at $\approx 0.04\ \mathrm{pc}$. Yet the central cusp is not located in a spherically symmetric configuration but in two young stellar disks within $0.5\ \mathrm{pc}$ of \mbox{Sgr~$\mathrm{A}^*$} containing a mass of $\sim 1.5\times 10^4\ M_\odot$.  

Further clues about the causes and effects of tidal evolution in a many-body system may come from detailed dynamical studies of the evolution of the Solar system, which have revealed a number of unexpected phenomena that perturb orbits  and tidally melt smaller bodies when they are caught in stochastic resonances \citep{1988Icar...76..295D,2005Icar..175..248F,2006ApJ...639..423M}.

As for the possible number of these satellites, we note that according to \citet{1995ApJ...455..342C} the Edgeworth-Kuiper belt of our Solar system may still (4.5 billion years after its creation) contain as many as $2\times 10^8$ objects of radii $\lesssim 10\ {\rm km}$.

On the basis of this discussion, one might expect to find stellar-system satellites all the way down to the black hole, a fair proportion of them on low-periastron, highly-eccentric orbits.

Close to low periastron, tidal forces apply significant work to objects. This work partially dissipates (in terms of heat, or accelerated electrons), lowering the orbital energy and initiating significant tidal evolution of the orbit, \v CCK08. Tidal interaction is strongest in resonance, i.e. when the orbital and fundamental quadrupole frequencies are identical. In this context, one must consider two classes of these satellites: those that are gravity-dominated and those that are solid-state-forces-dominated. For gravity-dominated objects, the fundamental frequency is $\nu_{\rm g} \approx 2 \sqrt{G \rho / 3\pi}$, while for solid-state-dominated objects, it is $\nu_{\rm s}=\frac{1}{4}c_{\rm s}/R$ ($\rho$ is the density of the body, $R$ its radius and $c_{\rm s}$ the speed of sound). If we assume typical values of $c_{\rm s}\approx 5\ \rm{km/s}$ and $\rho\approx 5\ \rm{g\ cm^{-3}}$, we ascertain the radius dividing the two classes to be about $1000\ \mathrm{km}$, i.e. the radius of the asteroid Ceres. Thus, all gravity-dominated satellites will have about the same fundamental quadrupole frequency, corresponding to the period of about 54 minutes. All smaller satellites will have shorter fundamental periods. Gravity-dominated satellites begin their rapid tidal evolution when their periastron reaches $r_{\rm p}\approx 10\ r_{\rm g}$ ($r_{\rm g} = GM_{\rm bh}/c^2$), while solid-state-dominated bodies may begin significant tidal evolution even closer to the black hole.

\begin{figure}
\resizebox{\hsize}{!}{\includegraphics{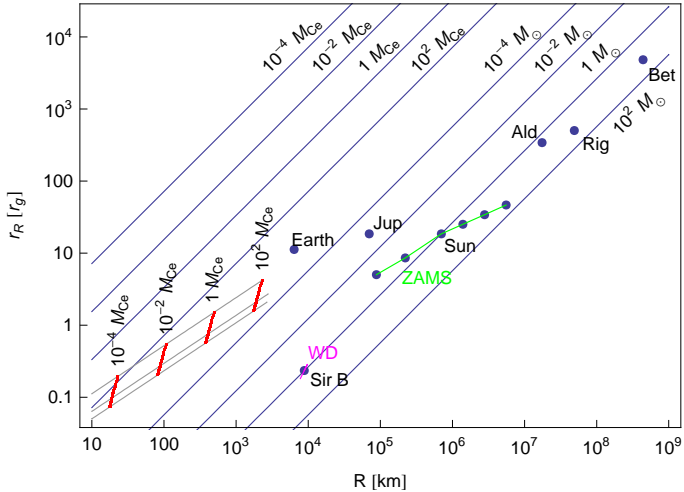}}
\caption{Roche radius and effective Roche radius as a function of mass and radius of the object. Blue lines indicate the Roche radius for gravity-dominated objects and red lines the effective Roche radius for solid-state-dominated objects ($M_{\mathrm{Ce}}$ is the mass of the asteroid Ceres). Lines of constant sound velocity $c_{\rm s}$ are plotted in light grey and refer to $c_{\rm s}=1.5,\ 3.5,\ 5\ {\rm km\ s^{-1}}$ (ice, rock, iron). Some celestial objects are marked by blue dots: Earth, Jupiter, Sun, Sirius B, Aldebaran, Betelgeuse and Rigel. The green line represents zero-age main-sequence stars and the purple line represents white dwarfs.}
\label{fig:RocheRadiji}
\end{figure}

To illustrate the difference between the two classes of objects, we introduce the Roche radius and the effective Roche radius. The effective Roche radius is defined as the radius at which  $\omega_{\rm q} T_{\rm f} = 1$, where $\omega_{\rm q}$ is the fundamental resonant frequency of quadrupole modes and $T_{\rm f}=(2 r_{\rm p}^3/ GM_{\rm bh})^{1/2}$ is the periastron crossing time \citep{2005ApJ...625..278G}.\footnote{The effective Roche radius becomes the Roche radius for gravity-dominated objects.} Figure\hspace{2pt}\ref{fig:RocheRadiji} shows the Roche radius and the effective Roche radius as a function of object size for different masses of gravity-dominated and solid-state-dominated objects. The positions of different celestial objects are also marked in this diagram. We note that the effective Roche radius for solid-state-dominated asteroids is below $2\ r_{\rm g}$, while the Roche radius for these objects is about $10\ r_{\rm g}$. 
\section{Tidal disruption of a small body and resulting light curves}
\label{sec:tidal}
We are interested in small objects orbiting the black hole on eccentric, low-periastron orbits, which are presumably brought there by processes described in Sect.\hspace{2pt}\ref{sec:small}. An object orbiting a black hole on such an orbit experiences high tides and is therefore heated at each periastron passage. As we showed in \v CCK08, for a range of object masses, solid-state forces, gravity, and heat conduction interact in an interesting way. Namely, the varying stress induced by the varying tidal field exerts work. This work is transferred into heat that escapes on a thermal diffusion timescale $\tau_{\rm solid}=R^2 \pi^{-2}D^{-1}$, of the order of $3\times 10^5$ years for 10 km objects consisting of normal, solid rock (thermal diffusion $D\sim 10^{-6}{\rm m}^2{\rm s}^{-1}$), which is long with respect to the orbital period. Thus, a stress that is well below breaking stress can do enough accumulated work  during the thermal diffusion time  to heat the body to melting temperatures. When melting begins, the solid-state structure weakens, allowing tides to do more work per passage, which leads to complete melting in a short time afterwards. In \v CCK08, we assumed that an object is on a sufficiently eccentric orbit ($e\gtrsim 0.1$) and that its fundamental resonance has the mechanical $Q$ value of 500. We found that objects with $R>20\ {\rm km}$ melt if $r_{\rm p}<5\ r_{\rm g}$ and those with $R>50\ {\rm km}$ melt if $r_{\rm p}<10\ r_{\rm g}$. However, heat produced by tides is insufficient to evaporate small bodies, because the weakening of solid-state structure allows convection to set in so that heat is efficiently brought to the surface, where it is radiated away at sub-evaporating temperatures. In this way, small solid-state-dominated objects that were orbiting the black hole above their effective Roche radius, find themselves below their effective Roche radius, simply because melting has moved their effective Roche radius closer to the Roche radius. This transition occurs without appreciable loss of orbital energy. It will be shown below that evaporation of the body occurs only during the last runs when the orbit develops into into an unstable, critical orbit.

In previous simulations, hot spots orbiting a black hole were treated either as point sources \citep{1992A&A...257..531K,  1994ApJ...425...63B, Schnittmann2005, 2006AN....327..957P} or small, fixed-size blobs \citep{1996ApJ...470..743K, 2004ApJ...606.1098S}. However, as we argued above, even solid objects larger than 10 or a few 10 km melt when their periastron is below $10\ r_{\rm g}$, and therefore rapidly deform afterwards. 

Here, we study the final stages of the tidal capture of such a small, melted object orbiting a Schwarzschild black hole.

To ensure that the calculations are manageable, we study the following simplified problem. We consider a solid, spherical object moving around a black hole on an elongated, bound orbit. After tidal forces do sufficient work, the object momentarily melts. The distance from the black hole at which this happens is denoted by $r_{\rm melt}$. After this moment, pressure forces become negligible with respect to inertial forces, and the object begins to behave as a collection of freely noninteracting particles. We use ray-tracing from points on the object's surface to the distant observer to determine its appearance and light curve. Ray-tracing procedures are described in \citet{2005PhRvD..72j4024C} and are freely available from the authors.

\subsection{Numerical code}
\label{sec:numerics}
The surface of the object is triangulated with respect to the mesh of spherical coordinates on the original sphere. Before melting, the sphere is considered to be rigid and non-rotating (zero-spin angular-momentum). Its centre of mass follows a time-like geodesic with a chosen specific energy $E/m$ and specific angular momentum $l/m$. The coordinates of mesh points are expressed in the Fermi-Walker transported frame local to the centre of mass and are initially constant with respect to proper time. The rigidity constraint prescribes the same velocity for each point on the sphere and thus different specific energy and angular momentum. When the sphere is melted, it finds itself below the Roche radius. From this moment on, the constraint no longer applies and surface mesh points begin to follow time-like geodesics specified with their inherited specific energy and angular momentum.

\begin{figure}
\resizebox{\hsize}{!}{\includegraphics{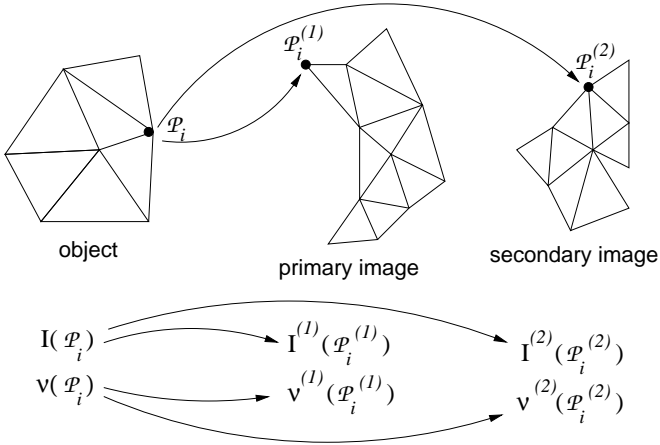}}
\caption{Mapping the triangulated sphere onto its image.}
\label{fig:preslikava}
\end{figure}

The object is considered as a monochromatic radiator (frequency $\nu_0$, constant rest-frame emissivity $I_0$) emitting according to Lambert's law in its own rest frame. The image of the object is  constructed by mapping triangles on the surface of the object onto the image plane of the observer and assigning the appropriate intensity and frequency to their image. As long as triangles on the object are sufficiently small and gravitational lensing is moderate, their images are also triangles that are defined by the positions of their vertices (see Fig.\hspace{2pt}\ref{fig:preslikava}). We note, however, that as time goes on, the stretching of the object increases exponentially, so that sooner or later every triangulation becomes too scarce. In such cases, we halt the simulation.
Through gravitational lensing, each vertex on the object produces multiple images. We note that the size of higher order images exponentially decreases with order, so it is sufficient to take into account only the primary and secondary image \citep{cadez3}.

\begin{figure}
\resizebox{\hsize}{!}{\includegraphics{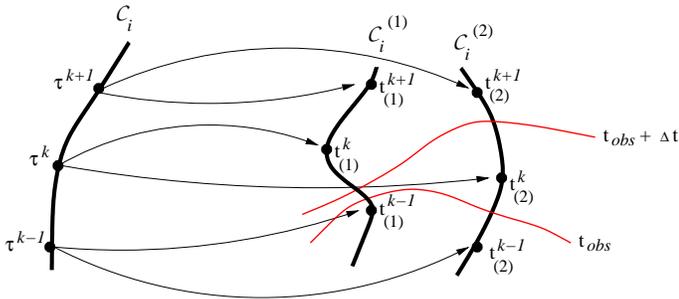}}
\caption{Determining the trails $\mathcal{C}_i^{(1)}$ and $\mathcal{C}_i^{(2)}$ of the image of a vertex $\mathcal{P}_i$ and its position at $t_{\rm obs}$ and $t_{\rm obs}+\Delta t$.}
\label{fig:interpolacija}
\end{figure}

The image is observed in the image plane at constant time intervals with respect to the observer. Since we are dealing with a highly dynamical problem, we must take into account that light travel times from the object to the observer are different for rays from different vertices. 
To calculate the position of the image of the vertex say ${\mathcal{P}_i}$ at observer time $t_{\rm obs}$, we first calculate the trail of the image of this vertex. We define $\mathcal{C}_i =\lbrace \tau, r_i (\tau),\theta_i(\tau),\phi_i(\tau)\rbrace$ to be the orbit of the point $\mathcal{P}_i$, parametrized with Schwarzschild coordinate time $\tau$. We send rays from $\mathcal{P}_i$ toward the observer and calculate the trails $\mathcal{C}^{(1)}_i=\lbrace t_i^{(1)}(\tau),x_i^{(1)}(\tau),y_i^{(1)}(\tau)\rbrace $ and $\mathcal{C}^{(2)}_i=\lbrace t_i^{(2)}(\tau),x_i^{(2)}(\tau),y_i^{(2)}(\tau)\rbrace $ of the primary and secondary image. Here $t_i^{(a)}(\tau)=\tau+\Delta t_i^{(a)}$ and $\Delta t_i^{(a)}$ is the time of flight from $\mathcal{P}_i$ to the observer along the $a$-th ray ($a= 1, 2$). The orbit is calculated at discrete time intervals $\tau^k$ and the coordinates in the image $\lbrace x_i,y_i\rbrace$ at time $t_{\rm obs}$ are calculated by linear interpolation as indicated in Fig.\hspace{2pt}\ref{fig:interpolacija}. The intensities $I_i^{(1)}$, $I_i^{(2)}$ and frequencies $\nu_i^{(1)}$, $\nu_i^{(2)}$ are interpolated in a similar way. 
In the image, all visible triangle areas $S_i$ are calculated and they contribute $\Delta L_i(t_{\rm obs})= I_i(t_{\rm obs})S_i(t_{\rm obs})$ to the total observed luminosity.

\subsection{Results}
\label{sec:results}
\begin{figure}
\resizebox{\hsize}{!}{\includegraphics{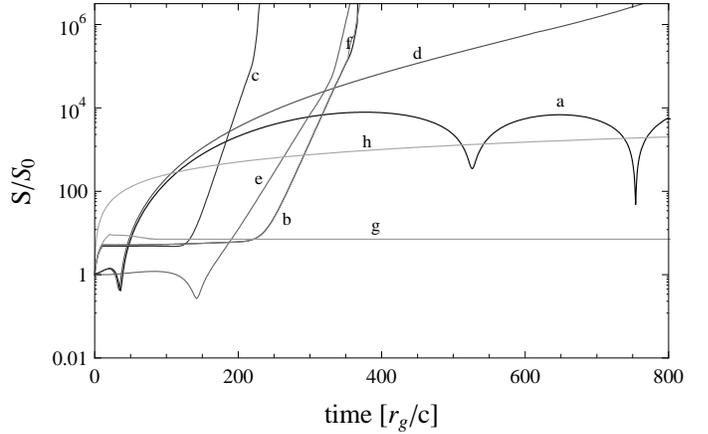}}
\caption{Area $S$ of the image of a tidally-evolving object seen above the orbital plane as a function of time for different orbital parameters: a) elliptic orbit close to innermost stable circular orbit (ISCO) with $r_{\rm A} = 6.2122\ r_{\rm g}$, $r_{\rm P}=6.2097\ r_{\rm g}$ and $\zeta=2.44\times 10^{-5}$; b) plunging parabolic orbit with $E/mc^2 = 1$ and $l/mr_{\rm g}= 3.999998$ and $\zeta=1+6.6\times 10^{-6}$; c) plunging hyperbolic orbit with $r(V_{\mathrm{max}}) = 3.6\ r_{\rm g}$ and $\zeta=1+10^{-6}$; d) elliptic orbit close to ISCO with $r_{\rm A} = 6.1\ r_{\rm g}$, $r_{\rm P} = 5.95\ r_{\rm g}$ and $\zeta=1-7\times 10^{-6}$; e) elliptic orbit with $r_{\rm A} = 20\ r_{\rm g}$, $r_{\rm P} = 4.45\ r_{\rm g}$ and $\zeta=1-7\times 10^{-6}$; f) scattering parabolic orbit with $E/mc^2 = 1$, $l/mr_{\rm g}= 4.000002$ and $\zeta=1-7\times 10^{-6}$; g) plunging hyperbolic orbit with $E/mc^2=3.0$, $l/mr_{\rm g}=6$, $\zeta=6.35$; h) linearly growing deformation. $S_0$ is the visible area of originally rigid sphere as seen by the observer. In cases b,c,e and f, the simulation was halted before the object deformed beyond a limit of numerical accuracy.}
\label{fig:ploscine}
\end{figure}

\begin{figure}
\resizebox{\hsize}{!}{\includegraphics{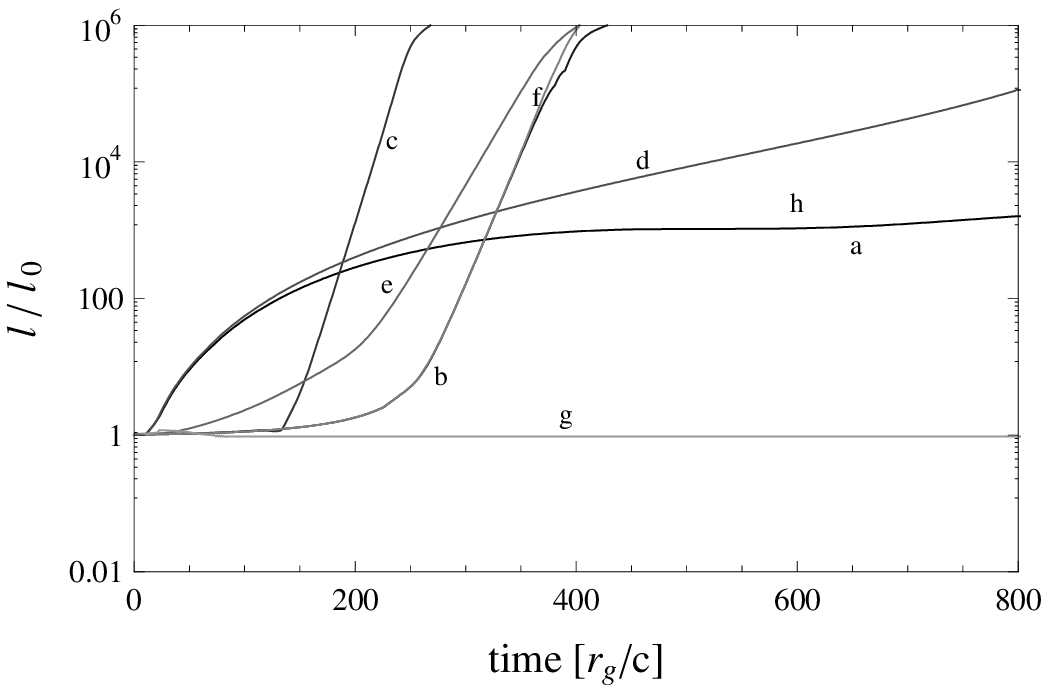}}
\caption{The length $l$ of the image of a tidally-evolving object seen above the orbital plane as a function of time for different orbital parameters: a) elliptic orbit close to innermost stable circular orbit (ISCO) with $r_{\rm A} = 6.2122\ r_{\rm g}$, $r_{\rm P}=6.2097\ r_{\rm g}$ and $\zeta=2.44\times 10^{-5}$; b) plunging parabolic orbit with $E/mc^2 = 1$, $l/mr_{\rm g}= 3.999998$ and $\zeta=1+10^{-6}$; c) plunging hyperbolic orbit with $r(V_{\mathrm{max}}) = 3.6\ r_{\rm g}$ and $\zeta=1+10^{-6}$; d) elliptic orbit close to ISCO with $r_{\rm A} = 6.1\ r_{\rm g}$, $r_{\rm P} = 5.95\ r_{\rm g}$ and $\zeta=1-3\times 10^{-6}$; e) elliptic orbit with $r_{\rm A} = 20\ r_{\rm g}$ , $r_{\rm P} = 4.45\ r_{\rm g}$ and $\zeta=1-3\times 10^{-6}$; f) scattering parabolic orbit with $E/mc^2 = 1$,  $l/mr_{\rm g}= 4.000002$ and $\zeta=1-3\times 10^{-6}$; g) plunging hyperbolic orbit with $E/mc^2=3.0$ and $l/mr_{\rm g}=6$ and $\zeta=6.35$; h) linearly growing deformation. In cases b,c,e and f, the simulation was halted before the object evolved beyond a limit of numerical accuracy.}
\label{fig:dolzine}
\end{figure}

\begin{figure}
\resizebox{\hsize}{!}{\includegraphics{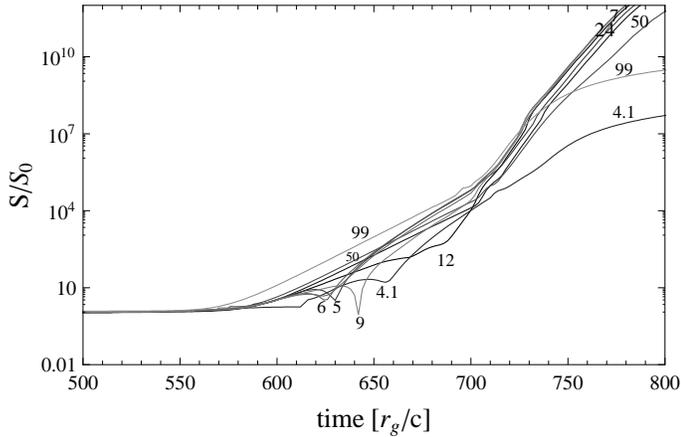}}
\caption{Area $S$ of the image of a tidally evolving object seen above the orbital plane as a function of time for $\zeta = 1-7\times 10^{-6}$ parabolic orbit (case `f' from Fig.\hspace{2pt}\ref{fig:ploscine}) with $r_0=100\ r_{\rm g}$ for different $r_{\rm melt}$. $S_0$ is the visible area of originally rigid sphere as seen by the observer. Note that the dependence on $r_{\rm melt}/r_g$ is weak}
\label{fig:ploscine1}
\end{figure}

We simulated objects moving along different orbits, characterized by different values of $E/m$ and $l/m$. We found that relativistic tidal effects depend predominantly on the ratio 
\begin{equation}
\zeta=(E/mc^2-V_{\rm min})/(V_{\rm max}-V_{\rm min})\ ,
\label{eq:zeta}
\end{equation}
Where $V_{\rm max}$ and $V_{\rm min}$ are the maximum and minimum of the relativistic effective potential $\widetilde V=\left((1-2r_{\rm g}/r)(1+(l/m c r)^2 )\right)^{1/2}$ \citep{1973grav.book.....M}. For stable circular orbits, $\zeta=0$. Tidal effects on these orbits are small and evolve as in the Keplerian case, where tidal deformations (below Roche limit) increase linearly. On the other hand, $\zeta \sim 1$ corresponds to semi-stable orbits\footnote{Such orbits partly wind around the black hole at radii in the interval $3r_{\rm g}<r<6r_{\rm g}$, make short excursions away from the black hole, and repeat the winding cycle or end up behind the horizon. They do not exist in the Keplerian case.} that extend to $V_{\rm max}$. Simulations exhibit exponentially growing tidal deformations (see Figs.\hspace{2pt}\ref{fig:ploscine} and \ref{fig:dolzine}), so that even small bodies, which generally suffer less, are eventually exponentially smeared into long, thin threads. The extent of tidal deformation is estimated from the length of the object's image as seen by the observer above the orbital plane. In our simulations, all the exponential timescales become similar, $\tau_c\approx 11\ r_{\rm g}/c$. To assess the dependence of the speed of tidal deformations on the initial melting radius $r_{\rm melt}$, we simulated motions along the $\zeta =(1-7\times 10^{-6})$ orbit. As shown in Fig.\hspace{2pt}\ref{fig:ploscine1}, tidal effects do not depend strongly on $r_{\rm melt}$. We note that the Keplerian case is quite different: if $r_{\rm melt}$ is close to periastron (below the Roche radius), tidal deformations increase linearly with time, while they do not increase at all if $r_{\rm melt}$ is close to apastron (above the Roche radius).

Here, we present light curves for two limiting cases: $\zeta \sim 1$ and $\zeta \rightarrow 0$ (see \citet{urosPhDThesis} for a detailed discussion). Time initiates from the moment of first detection by the observer.

\begin{figure}
\centering
\resizebox{0.5\hsize}{!}{\includegraphics{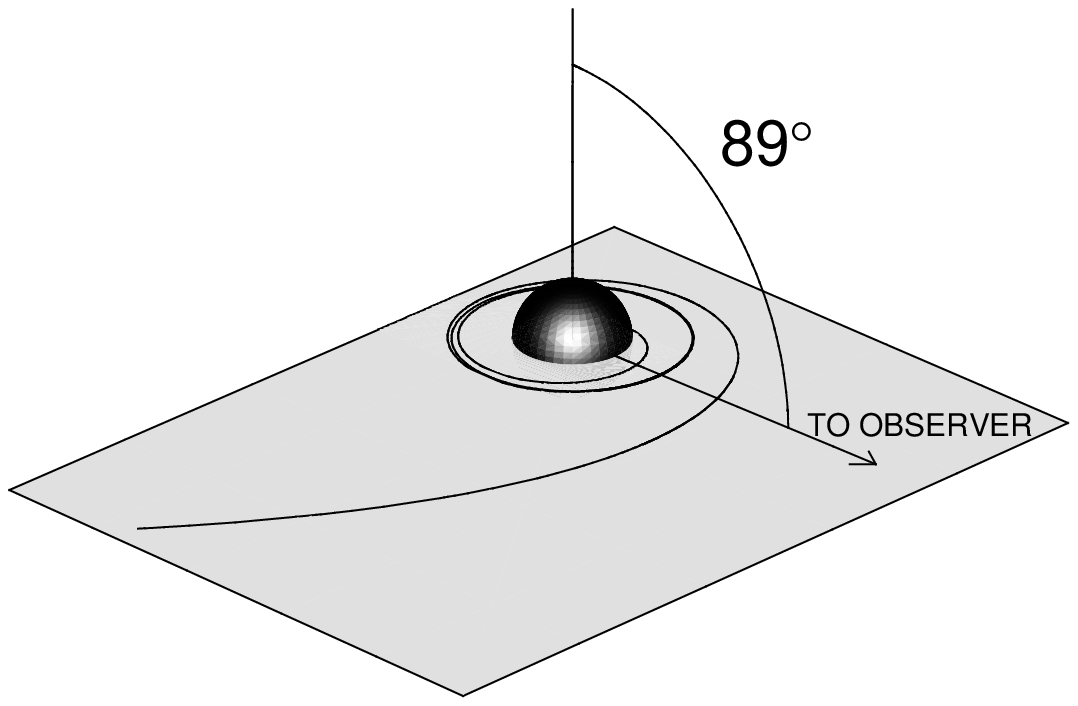}}
\resizebox{\hsize}{!}{\includegraphics{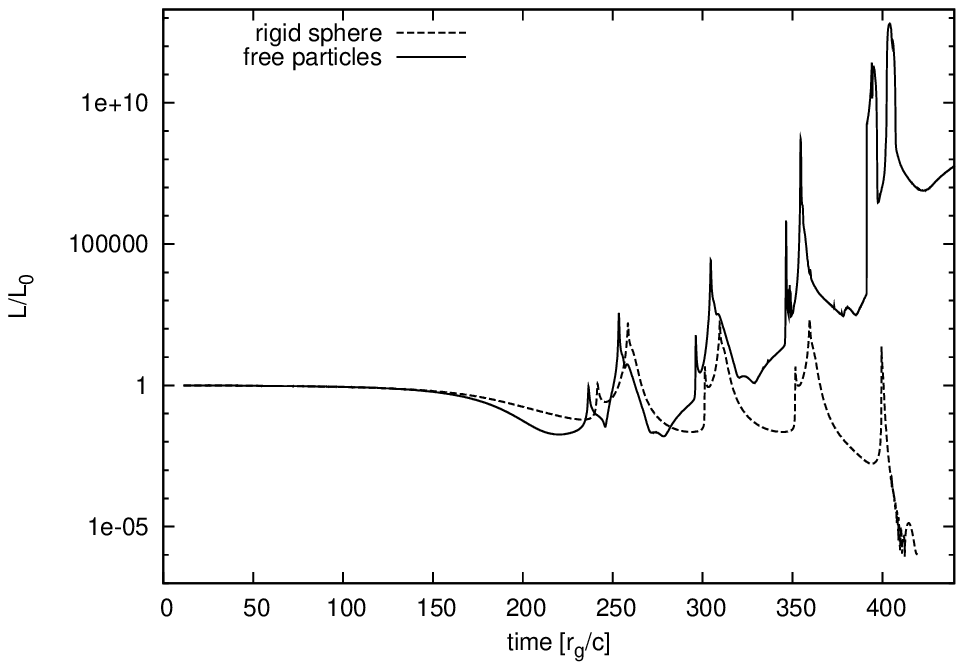}}
\caption{Top: Orbit of the object with $E/mc^2 = 1$ and $l/mr_{\rm g} c = 3.999998$. Bottom: The light curve of the object of size $R=10^{-5}\ r_{\rm g}$. The ``melting point'' is at $r_{\rm melt} = r_0 = 50\ r_{\rm g}$. The observer is $1\degr$ above the orbital plane. $L_0$ is the luminosity of originally rigid sphere.}
\label{fig:example}
\end{figure}

\begin{figure}
\resizebox{\hsize}{!}{\includegraphics{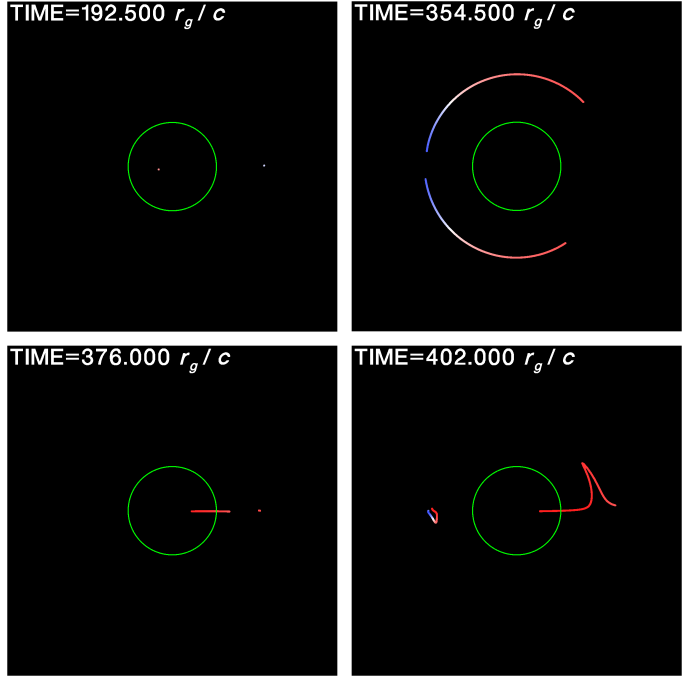}}
\caption{The image of a freely-falling surface that was initially spherical. It is seen by an observer $1\degr$ above the orbital plane. The colours on this tidally-distorted, gravitationally-lensed image correspond to redshift. The green circle represents the Schwarzschild radius.}
\label{fig:example_obj}
\end{figure}

\begin{figure}
\centering
\resizebox{0.5\hsize}{!}{\includegraphics{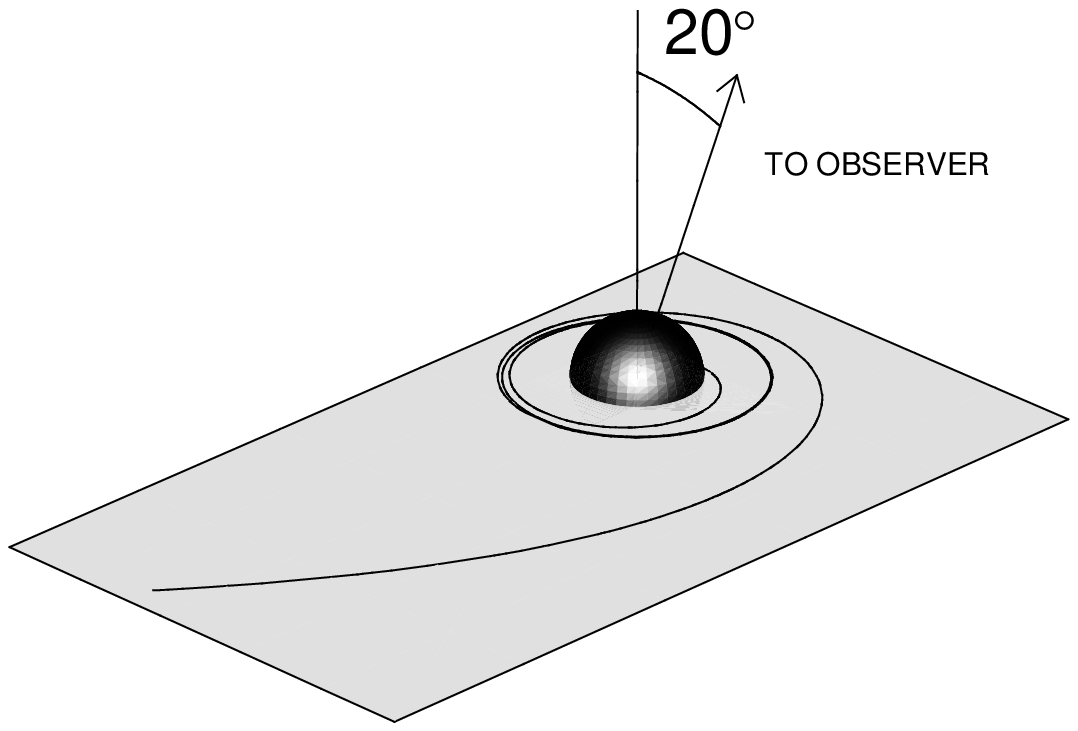}}
\resizebox{\hsize}{!}{\includegraphics{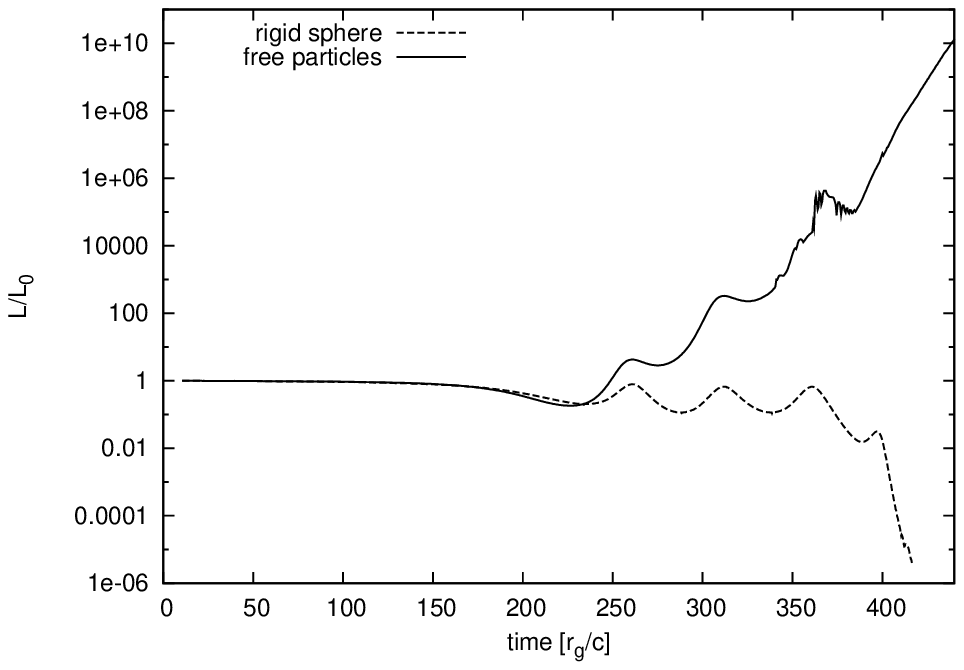}}
\caption{Top: Orbit of the object with $E/mc^2=1$ and $l/mr_{\rm g} c = 3.999998$.
Bottom: The light curve of the object of size $R=10^{-5}\ r_{\rm g}$. The ``melting point'' is at $r_{\rm melt} = r_0 = 50\ r_{\rm g}$. The observer is $70\degr$ above the orbital plane. $L_0$ is the luminosity of originally rigid sphere.}
\label{fig:example_20}
\end{figure}

\begin{figure}
\resizebox{\hsize}{!}{\includegraphics{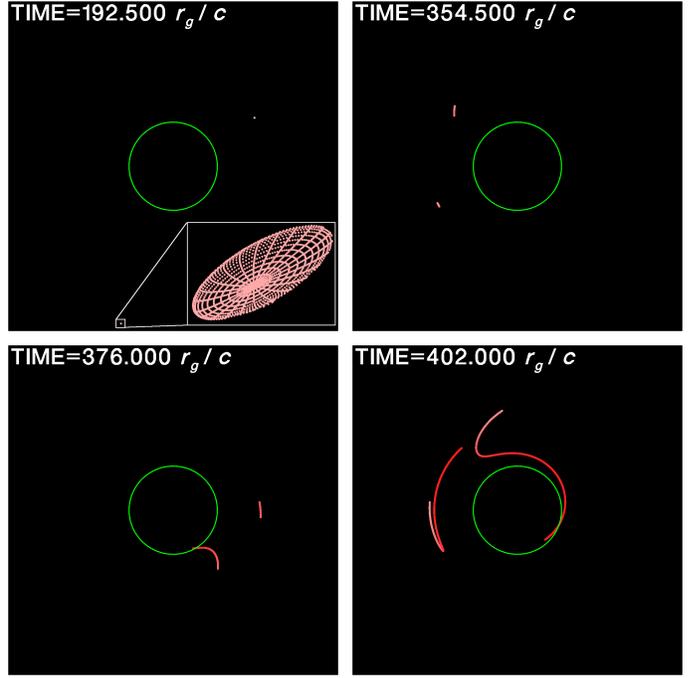}}
\caption{Freely-falling spherical surface, tidally-distorted seen through gravitational lensing, as seen by the observer $70\degr$ above the orbital plane. An enlargement of the primary image is included in the top left image. The colours correspond to the redshift. The green circle represents the Schwarzschild radius.}
\label{fig:example_obj_20}
\end{figure}

The $\zeta \sim 1$ case is represented by a parabolic orbit\footnote{A parabolic orbit in this context is meant as a synonym for highly eccentric elliptical orbit, while motion along hyperbolic orbits is considered as a test of proper behaviour of numerical models.} with $E/mc^2 = 1$ and $l/mr_{\rm g} c= 3.999998$ (case `b' in Fig.\hspace{2pt}\ref{fig:ploscine}). The starting point is at $r_0 = 50\ r_{\rm g}$ and $r_{\rm melt} = r_0$. Two examples of light curves are shown in Figs.~\ref{fig:example} and \ref{fig:example_20} with corresponding images of the object in Figs.~\ref{fig:example_obj} and \ref{fig:example_obj_20}. By comparing the top right images of Figs.~\ref{fig:example_obj} and \ref{fig:example_obj_20}, one realizes that in the case of Fig.\hspace{2pt}\ref{fig:example_obj}, where the observer is only $1\degr$ above the orbital plane, the small object has been gravitationally lensed into primary and secondary Einstein arcs. This magnification has a duration of about $1\ r_{\rm g}/c$. In the last two images of Fig.\hspace{2pt}\ref{fig:example_obj}, the object is already stretched into a thin thread as confirmed in Fig.\hspace{2pt}\ref{fig:example_obj_20}.

The sharp and narrow peaks in the light curves are due to gravitational lensing and always appear in pairs: the lower one is the contribution of the secondary image and precedes the one from the primary image. The peaks are higher if the observer is closer to the orbital plane (compare with Fig.\hspace{2pt}\ref{fig:example_20}). The wider bumps following sharp peaks are the result of Doppler boosting and the aberration beaming of light. The overall increase in luminosity is a consequence of the exponentially increasing surface of the object. 

To compare results with orbiting blobs of fixed size, we also show light curves of a freely falling rigid object in Figs.~\ref{fig:example} and \ref{fig:example_20}. These results are in agreement with \citet{1994ApJ...425...63B} and \citet{1996ApJ...470..743K}. Both light curves include the secondary image contribution.

\begin{figure}
\centering
\resizebox{0.5\hsize}{!}{\includegraphics{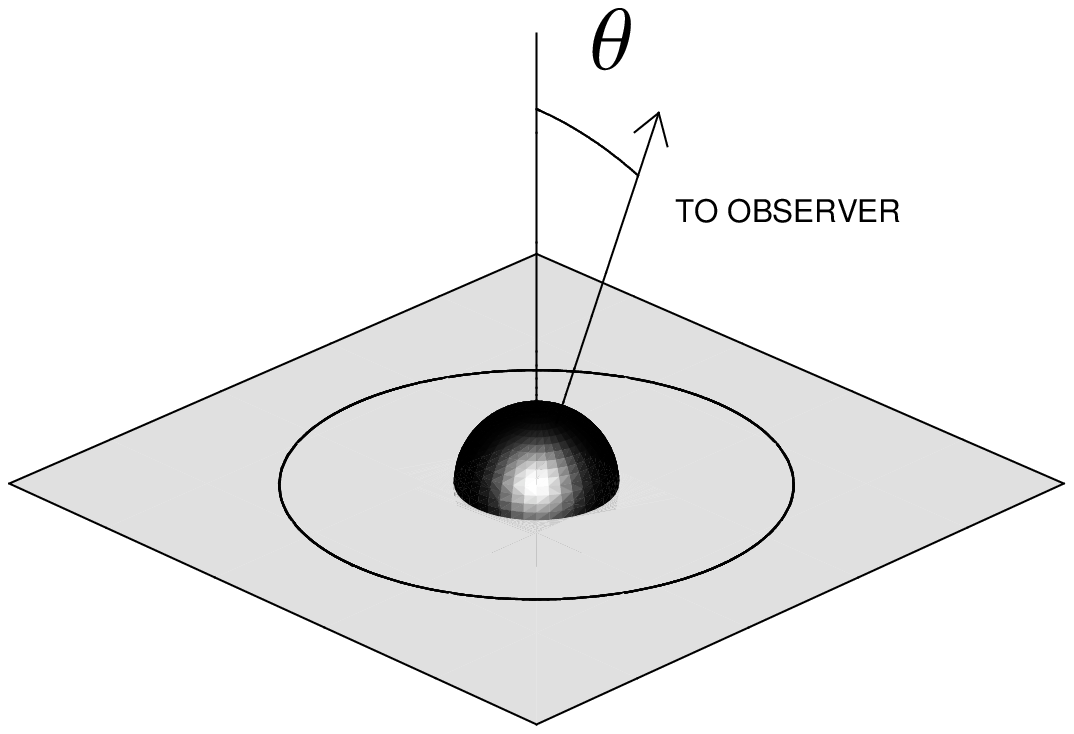}}
\resizebox{\hsize}{!}{\includegraphics{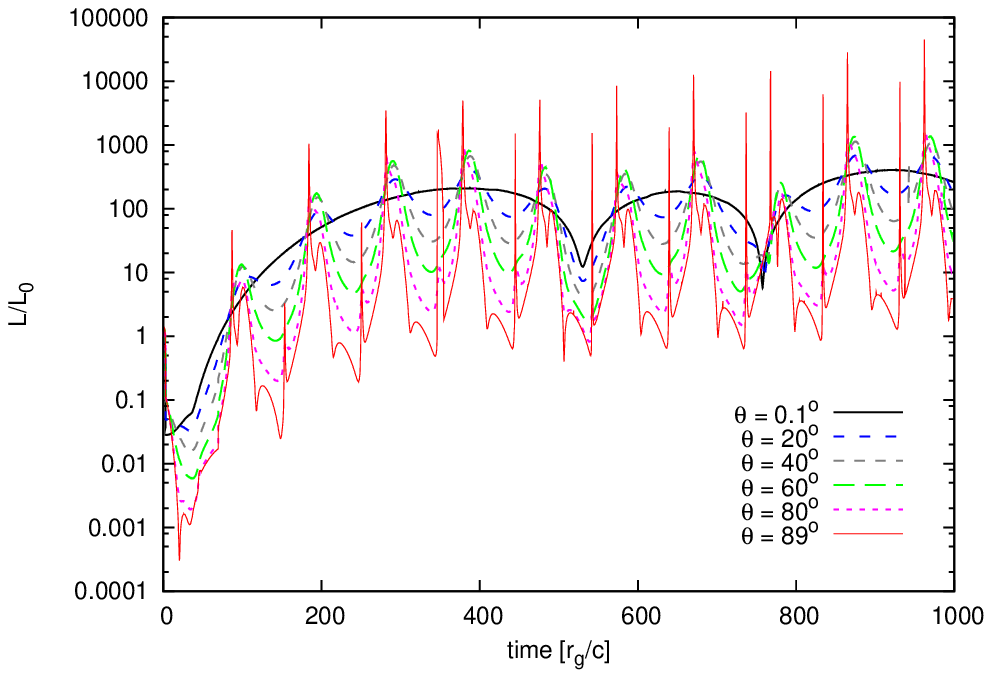}}
\caption{Top: Orbit of the object with $E/mc^2= 0.94294053$ and $l/mr_{\rm g} c = 3.46610164$. Bottom: Light curves of the object of size $R=10^{-5}\ r_{\rm g}$ for different angles $\theta$ of the observer. $L_0$ is the luminosity of originally rigid sphere.}
\label{fig:light_circular}
\end{figure}

\begin{figure}
\resizebox{\hsize}{!}{\includegraphics{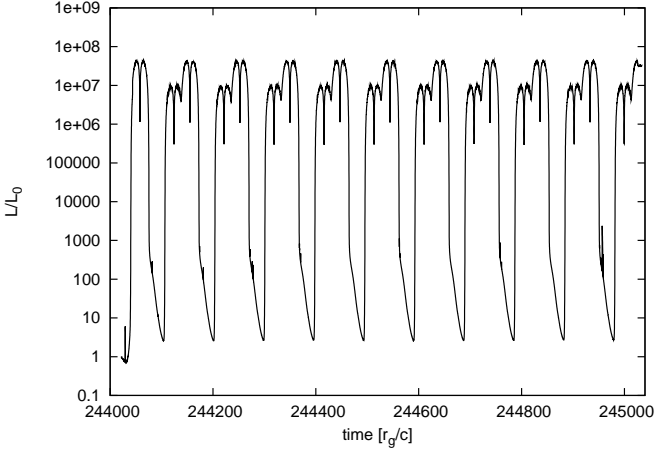}}
\caption{Same as Fig.\hspace{2pt}\ref{fig:light_circular} for $\theta = 89\degr$ of the observer but only at a much later time.}
\label{fig:light_last}
\end{figure}

The case $\zeta\rightarrow 0$ is illustrated by tidal evolution along a quasi-circular orbit with $E/mc^2 = 0.94294053$ and $l/mr_{\rm g} c = 3.46610164$, so that the apastron lies at $r_{\rm A}=6.216\ r_{\rm g}$ and periastron at $r_{\rm P}=6.206\ r_{\rm g}$ (case `a' in Fig.\hspace{2pt}\ref{fig:ploscine}). The starting point lies at $r_0 = 6.21\ r_{\rm g}$ and $r_{\rm melt} = r_{\rm A}$. The corresponding light curves for different observer's orientations are shown in Fig.\hspace{2pt}\ref{fig:light_circular}. In Fig.\hspace{2pt}\ref{fig:light_last}, we also show the light curve of the same object after a large number of periods. Since the light curve is not smeared out and extremely sharp spikes are still present at the end, we conclude that the object is not completely tidally disrupted; its size along the orbit increases (linearly) from $R=10^{-5}\ r_{\rm g}$ to e.g. $R \sim 0.1\ r_{\rm g}$.
\section{Discussion}
\label{sec:discussion}
\subsection{The effects of magnetic field}
\label{sec:magnetic}
Most authors now agree that radio and NIR flaring emission from \mbox{Sgr~$\mathrm{A}^*$} is produced by synchrotron radiation of electrons with Lorentz factors $\gamma_e \sim 10^2 - 10^3$ moving in a magnetic field ($10-100\ \mathrm{G}$) \citep{2001A&A...379L..13M, 2001Natur.413...45B, 2004ApJ...606..894Y}, while X-ray flares are assumed to be produced by inverse Compton scattering of low energy photons from relativistic electrons \citep{2004A&A...427....1E, 2006A&A...450..535E}. As to the observed magnetic field, there is still no general agreement on its strength at the center of the Galaxy. However, it has been pointed out by \citet{2006JPhCS..54...10L} that the dispute is whether the field is strong (of order $1\ \mathrm{mG}$) and globally organized \citep{1996ARA&A..34..645M, 2006JPhCS..54....1M}, or globally weak (of order $10\ \mathrm{\mu G}$) with regions in which it is stronger \citep{2005ApJ...626L..23L, 2006ApJ...637L.101B}. In any case the observed strength is orders of magnitude lower than requested by synchrotron models. Assuming that the object is permeated with such a low magnetic field, our simulations suggest a simple mechanism for increasing its strength as well as providing electrons at high $\gamma_e$.

Our simulations, discussed in the previous section, demonstrate that objects that find themselves on $\zeta\sim 1$ orbits become exponentially squeezed and elongated. Since the inward squeezing velocity is low ($v/c\sim 10^{-7}$), it is possible to use the MHD approximation for describing the evolution of the magnetic field
\begin{equation}
\frac{1}{\sigma \mu_0}\nabla\times(\nabla\times\vec{B}) =
-\frac{\partial \vec{B}}{\partial t} + \nabla\times(\vec{v}\times\vec{B})\ ,
\label{eq:induction}
\end{equation}
where $\sigma$ is the electrical conductivity and $\vec v$ is the plasma velocity. With typical values $\sigma=4.14 \times 10^6\ \Omega^{-1}\mathrm{m}^{-1}$ for molten aluminum \citep{conductivity} and $L=100\ \mathrm{km}$ for the size of the object, the magnetic Reynold's number $R_{\rm m}$ becomes large
\begin{displaymath}
R_{\rm m} = \mu_0 \sigma v L \approx 10^7\ .
\end{displaymath}
Consequently, the diffusion term of Eq.\hspace{2pt}\ref{eq:induction} becomes negligible, and we can write
\begin{equation}
\frac{\partial \vec{B}}{\partial t} = \nabla\times(\vec{v}\times\vec{B})\ ,
\label{eq:induction_frozen}
\end{equation}
which, according to Alfv\' en's frozen-flux theorem, leads to magnetic-flux conservation. As the object with initial magnetic field $B_0$ and cross section $S_0$ is exponentially squeezed, the magnetic field increases likewise as $B=B_0 S_0/S$. Taking into account that the tidal deformation tensor is traceless \citep{1973grav.book.....M}, the average cross-section area of a squeezed and elongated object is inversely proportional to its length, so that an approximate magnetic field at a later time can be estimated to be
\begin{equation}
B(t) \approx B_0 \frac{3}{4}\frac{r}{R_0} \Delta\varphi (t)\ ,
\label{eq:B}
\end{equation}
where $r$ is the distance from the black hole, $R_0$ the initial size of the object, and $\Delta\varphi$ the arc length of the object. The exponentially increasing value of $\Delta\varphi(t)$ can be either calculated from tidal acceleration (Sect.\hspace{2pt}\ref{sec:results}) or inferred directly from pictures, after taking into account relativistic length corrections.

At the beginning, we assume that the object is permeated with a (galactic) magnetic field $B_0=1\ \mathrm{mG}$, has a size of $R_0=50\ \mathrm{km}\approx 10^{-5}\ r_g$, later orbits around the black hole at $r=4\ r_g$ and is stretched into an arc with $\Delta\varphi=10\degr$. After inserting these numbers into Eq.\hspace{2pt}\ref{eq:B}, the estimated magnetic field can become as high as $B\approx50\ \mathrm{G}$. We also note that the initial magnetic field of the object may well be far higher than the assumed value of $1\ \mathrm{mG}$, since magnetic fields as high as Earth's magnetic field have been measured in asteroids \citep{1993Sci...261..331K}.

We also highlight that the exponentially-increasing magnetic field satisfies the conditions within a betatron. Therefore, by taking into account the initial magnetic field and the magnetic field rise-time, it is not difficult to obtain electrons of extremely high $\gamma$.

\subsection{Flares}
We consider whether the flare activity observed at the GC is due to the capture of a small body, of comparable mass to that of a comet or an asteroid. The only important physical constraints on the source are the requirement that: 1) the source can provide sufficient energy, and 2) it is  sufficiently small to produce observable modulation of the light curve.

The timescales in our simulations agree with the timescales of flares, if we assume that the source of the flares moves on an unstable, circular orbit at $r\lesssim 4\ r_{\rm g}$ around a Schwarzschild\footnote{All analyses has been completed in Schwarzshild space-time, but in order to fit light-curves it is sometimes preferable to consider Kerr space-time with a moderate value of the angular momentum parameter $a$.} black hole of mass $3.6 \times 10^6 M_\odot$. Since this is a $\zeta\sim 1$ orbit, the tidal work increases exponentially and the object is deformed exponentially on a timescale of $11.3\ r_{\rm g}/c$. In this case the timescale for magnetic diffusion is much larger than the dynamical timescale, therefore the magnetic flux is conserved. Consequently, as the object is squeezed, the magnetic field (exponentially) increases and the electrons with high $\gamma$ begin to emit synchrotron radiation.

We assume that the energy released during this process originates in the gravitational potential energy of the object and is thus proportional to its mass. Since potential energy differences in an orbit at $r\lesssim 4\ r_{\rm g}$ are of the order of a few percent of $mc^2$, it follows that objects producing them probably have masses of the order of $10^{20}\ \mathrm{g}$ \citep{2003Natur.425..934G}. If the sources of flares were gaseous blobs of such a large mass, they would find themselves below the Roche radius far away from the black hole, and would therefore be completely disrupted before producing any modulation of the light curve. We conclude that the source of the flares is a small, solid object that orbits the black hole above the effective Roche radius.

After many orbits, this object melts due to work exerted by tidal forces, thus its effective Roche radius increases quickly towards the Roche radius. As the melted object approaches the black hole on a $\zeta\sim 1$ orbit, it is deformed into a long, exponentially increasing, tidal tail (similar to tails in Figs.\hspace{2pt}\ref{fig:example_obj} and \ref{fig:example_obj_20}) as it falls within the Roche radius. Since the size of the object is initially small, it can still produce the modulation of the signal for a number of orbits, until it is completely smeared along the orbit.

Using ray-tracing techniques, the appearance and the luminosity of such an event as seen by observers was calculated at different inclinations and position angles with respect to the orbit of the infalling asteroid. In this simple toy-model, we introduced five parameters: mass of the black hole $M_{\rm bh}$, tidal heating timescale $\tau_{\rm h}$, length of the tidal tail $l_{\rm tail}$, inclination of the orbit $i$, and the longitude of the line of nodes $\Omega$. A robust description of the June 16th 2003 IR flare was found for the following parameter values: $M_{\rm bh}=4\times10^6 M_\odot$, $\tau_{\rm h}=2300\, {\rm s}$, $l_{\rm tail} = 10^8\, {\rm km}$, $i\approx 90^\circ$ \citep{2006AIPC..861..566C}. Despite the model simplicity, the value of $M_{\rm bh}$ that we find is in good agreement with the measured value.

\begin{figure}
\resizebox{\hsize}{!}{\includegraphics{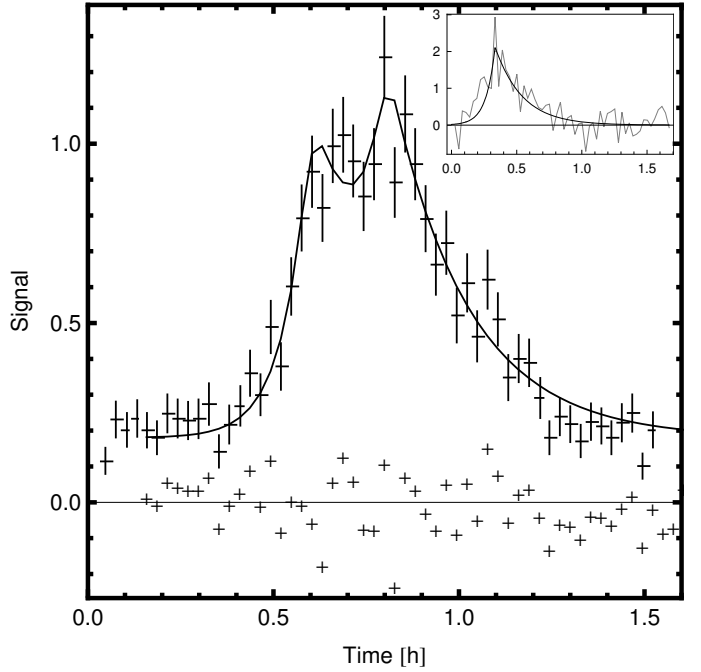}}
\caption{The XMM-Newton/EPIC light curve of the flare of April 4, 2007 \citep{2008A&A...488..549P}, fitted with the simple intensity distribution shown in the inset at top right. Residuals are shown as crosses at the bottom.}
\label{fig:fit}
\end{figure}

We were also able to describe the XMM-Newton/EPIC light curve of the flare of April 4, 2007 \citep{2008A&A...488..549P} with the following scenario: after the tidal deformation develops, the deformed object can be imagined as a long string of freely falling beads threading the same orbit. The light curve $S(t)$ of such a string can thus be described by a convolution
\begin{equation}
S(t)=\int G(t,t^\prime) \widetilde K(t^\prime) dt^\prime \ ,
\label{convol}
\end{equation}
where $G(t,t^\prime)$ is the Green's function belonging to the orbit, i.e. it is the signal, as a function of time $t$, of a single bead crossing a fixed reference point at time $t^\prime$, and $\widetilde K(t^\prime)$ is the distribution of luminosity among the beads. The Green's function depends on the orbital angular momentum $l$ and energy $E$ of the orbit. We are interested in plunging orbits with $\zeta\sim 1$, since tidal evolution produces these orbits, and only in such orbits can the length of the tidal tail grow exponentially and produce a flare. For such an orbit, the Green's function is dominated by a series of sharp spikes produced during the last few turns before plunging behind the horizon (see Fig.\hspace{2pt}\ref{fig:example}). The quasi-period of the spikes depends on the position of $V_{\rm max}$. We use the observed signal $S(t)$ and solve Eq.\hspace{2pt}\ref{convol} for the luminosity distribution $\widetilde K(t^\prime)$. Depending on the Green's function used, different solutions are obtained, most of which are physically unrealistic highly variable functions. Only for $l/m r_{\rm g}c = 4.3$ and $E = 1.04\ mc^2$ does the distribution become a simple positive function $\widetilde K(t^\prime)$, which can be well approximated with a steep, exponential rise, followed by a slower, exponential decay.\footnote{The energy and angular momentum were calculated by assuming that $M_{\rm bh}=3.6\times 10^6 M_\odot$.} Both $\widetilde K(t^\prime)$ and its smooth approximation  $K(t^\prime)$ are shown in the inset of Fig.\hspace{2pt}\ref{fig:fit}. 
The function $K(t^\prime)$ and the Green's function described above thus represent a viable description of a flare produced by an originally small object that is tidally distorted into a long thread during the last exponential part of its evolution in a $\zeta \approx 1$ plunging orbit. The exponential rise time of the flare is $4.8\ \mathrm{min}$ and the exponential decay time is $12.5\ \mathrm{min}$, which translates into a tidal-tail length of $l_{\rm tail} = 1.1\times 10^8\ \mathrm{km}$. The heating time that was estimated in the case of the IR flare cannot be estimated in this case, because we only have two peaks in the Green's function. In the case of the IR flare, the exponential rise time was considered as being rapid, and, considering the noise in the data, to be consistent with the short rise time preferred by the X-ray fit. The orbital energy obtained is probably unreasonably high. We note, however, that our analysis was performed in the Schwarzschild space-time. If this analysis were completed instead in Kerr space-time, the Green's function quasi-period would decrease for co-rotating orbits. For example, if the value of the Kerr parameter was $0.4$, the orbital energy would be only $\approx 0.945\ mc^2$.
\begin{figure}
\resizebox{\hsize}{!}{\includegraphics{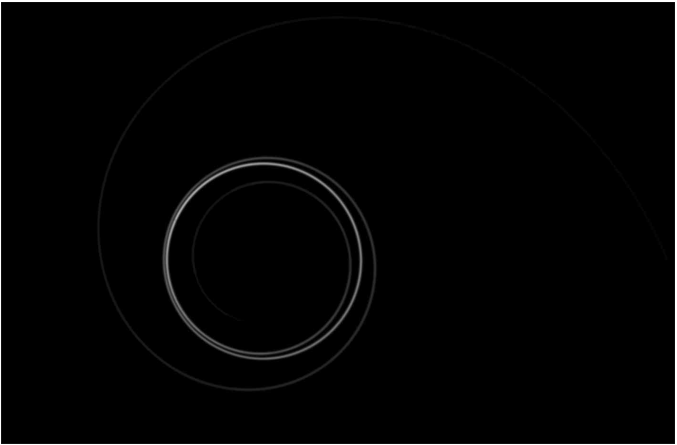}}
\caption{What the XMM-Newton/EPIC flare of April 4, 2007 might look like, if it could be seen from above the orbital plane.}
\label{fig:flare}
\end{figure}

On the other hand, if the object orbits the black hole on a $\zeta\sim 0$ orbit (i.e. ISCO), tidal work and deformation increase only linearly. In this case, the timescale for magnetic diffusion is much shorter than the dynamical timescale, the magnetic flux is therefore  not conserved and the magnetic field decays before reaching sufficiently high values to produce synchrotron radiation.

\section{Conclusions}
\label{sec:conclusions}
A puzzling characteristic of the Galactic centre that harbours a massive black hole are Galactic flares detected at various wavelengths. Their main characteristics are: a) short duration, b) quasiperiodic oscillations of period 17-22 minutes, c) total energy release of the order $10^{39.5}\ {\rm erg}$, and d) strong indications of the presence of magnetic fields of up to 100 gauss. 

The exact cause of the flares is still unclear, and several models have been proposed to explain their origin (see references in introduction). In this paper, we have proposed and explored the idea that flares are produced by the final accretion of dense objects with mass of the order of $10^{20}\ {\rm g}$ onto the massive black hole in SgrA$^*$. We have studied the effects of black-hole tides on small, solid objects in the vicinity of a massive black hole. Such objects are expected to populate the GC as the result of their being stripped from their parent stars. In \v CCK08, we showed that solid objects in the mass range $\sim 10^{19}-10^{21}\ {\rm g}$ melt in the near vicinity of the black hole. Afterwards their orbital evolution naturally leads to capture orbits, most likely by means of unstable, circular orbits. In an exhaustive numerical simulation, we studied the evolution of shapes of objects and their light curves as the objects move toward the event horizon of the black hole. We found that tides generally stretch the objects into long, thin threads that extend exponentially on a timescale $\tau_c\approx 11.3\ r_{\rm g}/c$, if the orbital criticality parameter $\zeta$ is close to 1. Furthermore, if the object is larger than $\gtrsim 0.3\ r_{\rm g}$, it is completely disrupted before producing any significant modulation of the light curve. Therefore, the object should be relatively small to produce the observed modulation.

Tidal evolution of the shapes and the images of these objects was then studied numerically in the space-time of a Schwarzschild black hole. Because of the complexity of the problem, we did not consider the more general Kerr space-time and instead limited the analysis to pressureless gas. We found in general, that these objects are eventually elongated into thin threads that in $\zeta\sim 1$ orbits extend exponentially.

During this exponential elongation, we highlighted that conditions for magnetic-flux conservation were obeyed, so that relatively high magnetic-field strengths could be generated. The exponential growth of the magnetic-field density also allows the betatron mechanism to operate, and possibly generate highly relativistic electrons. We propose this process as a possible source of Galactic flares. We have not yet developed the precise mechanism for extracting radiation from gravitational potential energy, but we are able to model light curves of Galactic flares with simple assumptions.

Finally, we emphasize that our scenario differs from other models in the following respects: 1) Flares originate in marginally-bound and not marginally-stable orbits. This allows them to travel closer to the black hole and provides a mechanism for the rapid extraction of energy. It also agrees with measurements of the size of the flaring region in the Galactic centre, which imply that it is not larger than $6\ r_{\rm g}$ \citep{2008arXiv0808.2624R}, and relaxes constraints on the angular momentum of the black hole. 2) Our model does not require a fairly rapidly rotating black hole as needed by \citet{2003Natur.425..934G}, \citet{2004A&A...417...71A} and \citet{2007MNRAS.375..764T}. 3) An accretion disk at the Galactic centre is not required. \citet{2008ApJ...682..361Y} highlited that ``there is no definitive evidence that \mbox{Sgr~$\mathrm{A}^*$}\ has a disk''.  4) Elongated structures invoked in other models to fit flare data (e.g. \citet{2007ApJ...662L..15F,2008JPhCS.131a2008Z} and \citet{2008arXiv0810.4947H}) develop naturally in the proposed model as a result of the tidal evolution of a melted body (see Fig.\ref{fig:flare}). 5) The evolving elongated structures also provide a natural explanation of rapid changes in magnetic-field density inferred from observations, increasing first as the body is squeezed by tides and finally decreasing as the body crosses the horizon of the black hole.

\begin{acknowledgements}
U.~K., A.~\v C. and M.~C. acknowledge support from the bilateral protocol of scientific and technological cooperation between Italy and Slovenia. U.~K., A.~\v C. and A.~G. acknowledge partial support from the grant 1554-501 of the Slovenian Research Foundation.
\end{acknowledgements}
\bibliographystyle{aa}
\bibliography{biblio}
\end{document}